# Radiation-Tolerant, High-speed Serial Link Design with SRAM-based FPGAs

R. Giordano, S. Perrella, and D. Barbieri

*Abstract*—High-speed serial links implemented in SRAM-based FPGAs have been extensively used in the trigger and data acquisition systems of High Energy Physics experiments. Usually, their application has been restricted to off-detector, mostly due the sensitivity of SRAM-based FPGA to radiation faults (single event upsets). However, the device tolerance to radiation environments can be achieved by adopting dedicated mitigation techniques such as information redundancy, hardware redundancy and configuration scrubbing.

In this work, we discuss the design of a bi-directional serial link running at 6.25 Gbps based on a Xilinx Kintex-7 FPGA. The link is protected against single event upsets by means of all the above-mentioned methods.

A self-synchronizing scrambler is used for DC-balance and data randomization, while the subsequent Reed-Solomon encoder/decoder detects and corrects bursts of errors in the transmitted data. The error correction capability of the line code is further increased by adopting the interleaving technique. Besides, in order to completely take advantage of available bandwidth and to cope with different rates of radiation-induced faults, the link can modulate the protection level of the Reed-Solomon code. The reliability of the link is also improved by means of modular redundancy on the frame alignment block. Besides, on the same FPGA, a scrubber repairs corrupted configuration frames in real-time.

We present the test results carried out using the fault injection method. We show the performance of the link in terms of mean time between failures (MTBF) and fault tolerance to upsets.

*Index Terms*—FPGA, Link, Adaptive, Tolerance.

## I. Introduction

HIGH-SPEED serial links implemented within Static-RAM based FPGA (SRAM-based FPGAs) have been extensively used in the trigger and data acquisition (TDAQ) systems [1-3] of High Energy Physics (HEP) experiments. However, their application has been usually restricted to off-detector, mostly due SRAM-based FPGA sensitivity to radiation faults. Instead, rad hard ASIC counterparts have been generally adopted [4] on-detector. Despite of this limitation, SRAM-based FPGAs flexibility, high performance and reduced design costs are suitable features which led to the development of mitigation techniques [5-6] to guarantee FPGAs reliability also in harsh radiation environments.

In order to cope with faults occurring in combinational and sequential logic, hardware redundancy strategies have been elaborated while configuration scrubbing techniques have been developed to take care of memory configuration faults. Finally, information redundancy methods have been implemented for preventing failures in data transmission.

For SRAM-based FPGAs, memory scrubbing is necessary to allow fault tolerance and prevents failures due to error accumulation. It has become even more important with newer FPGA families in which the increase in memory configuration size has led to a corresponding increase in cross-section failure overall. Configuration scrubbing [7] technique repairs a fault bit in the configuration memory content by reloading its correct value, without interrupting functionality of the device. To date, various scrubbing schemes for FPGAs have been proposed. Some of them foresee a periodic and complete rewrite of configuration memory independently from the faults presence (blind scrubbing). However, especially in the last few years, configuration read back technique has become one of the most widely used solutions. In this case, a reference copy of the configuration memory is compared with the current internal memory, directly read back from the FPGA. When a discrepancy is found, only the erroneous configuration frame is overwritten with the correct values obtained by the golden copy.

Information redundancy is used to guarantee a higher reliability in data transmission. This kind of redundancy is used by all Error Checking and Correcting (ECC) codes, such as Reed Solomon code [8]. It consists of adding extra information to the user data. In such a way, it is possible to detect and eventually correct faults without interfering in the system operations.

Hardware redundancy is another widely used technique for improving FPGA reliability. In this approach, functional blocks are replicated and usually placed in separate locations inside the device. Typically, critical blocks are implemented in three (Triple Modular Redundancy, TMR) or two (Dual Modular Redundancy, DMR) copies working in parallel and a final voter is used to detect errors. In the case of TMR, it can also determine the correct result. Indeed, if one of the replica fails, the majority voter hides the fault by using the output of

Manuscript received June 24, 2018.

S. Perrella is with "Università di Napoli Federico II, Dipartimento di Fisica" and I.N.F.N. sez. di Napoli, Complesso Universitario di Monte S. Angelo, via Cintia, 80126, Napoli – Italy (e-mail: sabrina.perrella@na.infn.it).

R. Giordano, and D. Barbieri are "Università di Napoli Federico II, Dipartimento di Fisica" and I.N.F.N. sez. di Napoli, Complesso Universitario di Monte S. Angelo, via Cintia, 80126, Napoli - Italy.

This work is part of the ROAL project (grant no. RBSI14JOUV) funded by the Scientific Independence of Young Researchers (SIR) 2014 program of the Italian Ministry of Education, University and Research (MIUR).

the two fault-free modules. As result, even if TMR is more expensive in terms of area required for implementation with respect to DMR, it is often preferred for its capability of reducing the frequency of faults. Depending on the goal to be achieved in the design and the radiation environment, a combination of different fault mitigation techniques can be adopted.

In this work we discussed the bi-directional serial link running at 6.25 Gbps implemented in the Xilinx Kintex-7 [9] SRAM-FPGA which could be used in radiation areas. For this reason, it has been protected against radiation faults by means ECC code, TMR and configuration scrubbing. The link adopts a line code in which a self-synchronizing scrambler is used for data randomization, a Reed-Solomon encoder/decoder detects and corrects errors and an interleaver doubles the error correction capability. The Reed-Solomon code adds an overhead which could be unacceptable in some applications. For this reason, we implemented a serial link capable of adjusting its Reed-Solomon coding, according to the different radiation environments and then the amount of redundancy in the serial stream. Thus, our link is capable of coping with the radiation induced faults, but at the same time to fully benefit from the available link bandwidth.

The reliability of the link is also improved by means of TMR. Synopsys Synplify Premier tool [10] has been used to implement distributed TMR on the critical block which aligns the data stream to the packet boundaries. Distributed TMR approach triplicates the sequential logic and the combinational logic within a module, including the voter logic on the sequential loops and module outputs.

Besides, in order to protect the configuration memory from radiation faults, on the same FPGA a scrubber continuously reads back memory contents and eventually repairs the corrupted bits. In our work the scrubber is an integral part of the link and it implements in hardware the scrubbing technique described in [11].

We developed an original software fault-injector for corrupting specific portions of the configuration memory device [12] and of the serial stream, and a tester circuit for monitoring the link response to the injected faults.

We presented the test results carried out using the fault injection method in terms of mean time between failure (MTBF) for its subcomponent. We also present measurements of mean time between loss of lock as characterization of the link.

## II. Conclusion

The 6.25 Gbps serial link we implemented in the Xilinx Kintex-7 [9] SRAM-FPGA, is able to correct burst errors in the serial stream. Besides, due to the TMR and the scrubber working on the same FPGA, the link can triplicate the tolerance to induced faults.